# Indication of Stochastic Photothermal Dynamics around a Topological Defect in a Chiral Magnet

Dongxue Han[1,2], Asuka Nakamura[1,2,*], Takahiro Shimojima[1,†], Kosuke Karube[1], Yasujiro Taguchi[1,3], Yoshinori Tokura[1,2,4], and Kyoko Ishizaka[1,2]

[1]*RIKEN Center for Emergent Matter Science (CEMS), Saitama 351-0198, Japan*

[2]*Quantum-Phase Electronics Center and Department of Applied Physics, The University of Tokyo, Tokyo 113-8656, Japan*

[3]*RIKEN Baton Zone Program, TRIP Headquarters, Saitama 351-0198, Japan*

[4]*Tokyo College, The University of Tokyo, Tokyo 113-8656, Japan*

[*]Author to whom correspondence should be addressed: asuka.nakamura@riken.jp

[†]Present address: *Department of Physics, Nagoya University, Nagoya 464-8601, Japan*

## Abstract

Chiral magnets host topologically protected spin textures whose nonequilibrium dynamics are crucial in phase transitions and domain evolution, yet ultrafast defect-mediated processes remain poorly understood. Here, we investigate photothermally induced helical-to-paramagnetic phase transition in $Co_9Zn_9Mn_2$ using pump-probe Lorentz transmission electron microscopy (LTEM). Following the suppression of the magnetic stripe contrast induced by femtosecond pulsed laser, we observe a directional recovery process of magnetic order driven by the anisotropic thermal diffusion, toward the thick region that effectively acts as a heat sink. Remarkably, around a magnetic edge dislocation, the magnetic contrast recovery exhibits a pronounced delay accompanied by a transient blurring of LTEM contrast. These findings suggest that the recovery dynamics around the magnetic edge dislocation proceed through multiple relaxation paths that are selected stochastically. Our results indicate a possible enhancement of stochasticity around topological defects during the recovery dynamics of magnetic phase transitions.

**Text**

Chiral magnets host a rich variety of topological spin textures, including domain walls, skyrmions, and edge dislocations, emerging from the competition between ferromagnetic exchange and the Dzyaloshinskii-Moriya interaction (DMI) associated with their broken inversion symmetry[1–3]. Representative chiral magnets such as FeGe, MnSi, and $Co_9Zn_9Mn_2$ stabilize helical, conical, and skyrmion lattice phases, whose characteristic length scales are determined by the ratio of exchange stiffness to DMI strength[4–7]. The dynamics of topological spin textures in these systems under external stimuli, such as magnetic fields, electric currents, or optical excitation, have been intensively investigated, revealing novel phenomena arising from their nontrivial topology[8,9]. Recently, photoinduced dynamics in chiral magnets have attracted growing interest as a promising route to achieve ultrafast control of magnetism on picosecond-to-nanosecond timescales[10,11]. Previous studies have revealed various photoinduced phenomena in chiral magnets, including helical-to-conical transitions[12], coherent domain wall propagation[13], and the creation, annihilation, and collective excitation of magnetic skyrmions[14,15].

Topological defects, such as edge dislocations and Bloch points, are expected to play a central role in the nonequilibrium dynamics of chiral magnets, particularly during magnetic phase transitions, where their nucleation, motion, and annihilation can govern domain growth and relaxation kinetics in general[16]. Defect-mediated dynamics have indeed been reported in a range of phenomena, including skyrmion annihilation[17,18], magnetic edge dislocation propagation[5,19], and helical phase reorientation[20,21]. Although the nanoscale observation of the spin textures and their dynamics has been achieved using several imaging techniques, including magnetic force microscopy[5], X-ray microscopy[22], and Lorentz transmission electron microscopy (LTEM)[23], direct and microscopic investigation of topological defects during photoinduced magnetic phase transitions has long been limited by the lack of temporal resolution. In this context, the recent integration of the pump-probe method using a femtosecond laser has further enabled nanoscale magnetic imaging with high temporal resolution[24–27], providing a powerful tool for investigating photoinduced dynamics associated with topological defects.

In this study, we investigate the photothermally induced helical-to-paramagnetic phase transition in a chiral magnet $Co_9Zn_9Mn_2$ using time-resolved LTEM measurements. This approach enables direct visualization of the photothermally induced magnetic phase transition and its subsequent recovery to the initial state with nanometer-nanosecond precision. The experimental results reveal a directional recovery of the magnetic stripe contrast driven by the anisotropic thermal diffusion toward the thick region that acts as an effective heat sink. Fourier transform analysis of the transient magnetic contrast images uncovers a non-monotonic spatial dependence of the relaxation time, which reaches its maximum around a magnetic edge dislocation. In addition, we observe a transient change in LTEM contrast, with significant blurring and a change in the magnetic pattern around the magnetic edge dislocation. Since multiple stochastic processes can be averaged in pump-probe measurements, this behavior can be attributed to an average of several different relaxation paths involving the slipping motion of the magnetic edge dislocation. These findings demonstrate the unique nonequilibrium magnetic dynamics that can emerge around topological defects during photothermally induced phase transitions in chiral magnets.

Bulk $Co_9Zn_9Mn_2$ crystals were synthesized following the procedures described elsewhere[28,29]. The thin plate sample was fabricated using a focused ion beam (Helios 450, Thermo Fisher Scientific), following standard TEM sample preparation procedures. A $Co_9Zn_9Mn_2$ lamella (> 500 nm thick) was first prepared, and then an 8 μm × 2.5 μm region was further thinned to 120 nm for TEM observation. Time-resolved LTEM measurements were performed on this thin plate region (~ 120 nm) adjacent to the thick region (> 500 nm) of the sample.

Pump-probe LTEM measurements were conducted using a previously described setup[27,30] based on Tecnai Femto (Thermo Fisher Scientific). A 150 μm condenser aperture was used for all measurements. For the pump laser, we used PHAROS (Light Conversion) with a pulse duration of 290 fs and a repetition rate of 25 kHz. The fundamental 1030 nm light is converted to 531 nm using an optical parametric amplifier, ORPHEUS (Light Conversion), to excite the sample. The pump beam diameter (~ 300 μm) is sufficiently larger than the sample size (< 10 μm), ensuring homogeneous excitation of the sample at a fluence of 5.0 mJ/cm$^2$. For probe electron generation, 10 ns, 266 nm optical pulses are delivered by AWave-532 (Advanced Optowave) and irradiate a carbon-coated $LaB_6$ photocathode (Applied Physics Technologies) mounted in the TEM. The total time resolution is approximately 10 ns, which is primarily governed by the pulse duration of the probe electron. Fresnel LTEM measurements are performed at a defocus value of $\Delta f = 1$ mm, with a spatial resolution of better than 50 nm. The acquisition time for each time delay is 20 minutes. The sample is tilted by 9.6° under an out-of-plane magnetic field of 50 mT to align the helical stripes. All data analysis and visualization were performed using the multi-dimensional data analysis platform *lys*[31].

We use $Co_9Zn_9Mn_2$ as a platform to investigate photothermally induced magnetization dynamics around a magnetic edge dislocation. $Co_9Zn_9Mn_2$ is a representative *β*-Mn-type chiral magnet with a non-centrosymmetric cubic structure belonging to space group $P4_132$ or $P4_332$ [Figure 1(a)]. This crystal symmetry gives rise to DMI, which stabilizes a helical magnetic state at room temperature[1]. Figure 1(b) shows a schematic magnetic phase diagram of the present sample. Previous studies have reported that $Co_9Zn_9Mn_2$ exhibits a paramagnetic phase above the Curie temperature $T_C \approx 390$ K, whereas near room temperature it transitions among helical, conical, and ferromagnetic phases depending on the external magnetic field[7,28]. In our fabricated thin plate, we did not confirm skyrmion lattice phase. Temperature-dependent LTEM measurements verify $T_C \approx 380$ K under static conditions: the helical phase displays periodic stripe contrast, while the paramagnetic phase shows no magnetic contrast [Figure 1(c)]. The helical period, determined by the ratio of exchange stiffness to DMI strength, is approximately 150 nm. We also confirm the helical-to-conical phase transition at room temperature under an external magnetic field of 0.1 T. We performed time-resolved LTEM measurements in the solid red rectangle in Figure 1(d), where the helical-to-paramagnetic phase transition is observed. Since the thickness of the thick region (> 500 nm) far exceeds the optical penetration length of the pump laser (typically < 100 nm) in metallic materials, only the thin plate region is effectively excited. As a result, we can track the recovery process of the helical state that is driven by the thermal diffusion near the boundary of thin/thick regions [Figure 1(e)].

Figure 2(a) shows representative LTEM images before and after the pulsed laser excitation. The images (except for $t = 20$ ns) are averaged over a $\pm$ 100 ns time window around the displayed time delay values to suppress diffraction contrast and improve the S/N ratio. At $t < 0$, a clear periodic stripe contrast of the helical state is visible, whereas at $t = 20$ ns, the stripe contrast completely disappears except near the thin/thick boundary. This rapid loss of contrast indicates the photothermal suppression of the helical magnetization, i.e., a photoinduced phase transition to the paramagnetic phase. Near the thin/thick boundary, the stripe contrast remains quite robust, indicating that the temperature in this region is kept relatively low compared with the central regions due to rapid thermal diffusion into the thick region. On longer timescales, the stripe contrast gradually reappears from the left edge toward the right side, and by $t \sim 1500$ ns the stripe pattern nearly recovers its original helical state. This spatially directional process can be attributed to the magnetization recovery driven by thermal diffusion. Figure 2(b) schematically illustrates laser-induced temperature evolution in the observed region. Immediately after excitation, only the thin region is heated above $T_C$ since the thickness of the thick region (> 500 nm) far exceeds the optical penetration depth of metallic $Co_9Zn_9Mn_2$ (< 100 nm). As a result, a strong in-plane temperature gradient drives heat flow toward the thick region, causing the thin plate region to cool back below $T_C$. In addition to this directional magnetization recovery, the magnetic pattern around the edge dislocation enclosed by the solid frame in Figure 2(a) shows distinctive transient behavior. Although by $t \sim 1500$ ns the LTEM image around the edge dislocation is very similar to that at $t < 0$, significant blurring appears at $t = 800$ ns, as highlighted by the solid rectangles in Figure 2(a). The spatial extent of the blurring is limited to a few times the helical period, as no blurring is observed in the bottom region. This blurring implies a transient change in the magnetic pattern within this temporal range, which will be discussed later.

To quantify magnetization recovery around the magnetic edge dislocation, we analyze the local stripe contrast in a selected region shown in Figure 3(a). Figure 3(b) is the fast Fourier transform (FFT) of the LTEM intensity along the $y$ direction at a representative $x$ position ($x = 900$ nm), which shows a clear peak at the helical wavevector. By integrating the $x$-dependent FFT intensity in the unshaded region in Figure 3(b), we quantitatively evaluate the local stripe contrast ($I_{FFT}$) as a function of the distance from the left edge ($x$). It should be noted that $I_{FFT}$ reflects both the magnitude of magnetization and changes in the magnetic pattern in real space. Figure 3(c) shows the time evolution of $I_{FFT}$ at representative $x$ positions, where each trace is averaged over a ± 100 nm $x$ window to improve the S/N ratio. While $I_{FFT}$ at $x = 300$ nm shows a simple relaxation toward the initial value ($t < 0$), $I_{FFT}$ at $x = 1100$ nm shows a pronounced plateau of about 500 ns during which $I_{FFT}$ remains nearly constant. This plateau can be interpreted as a temporal interval in which the magnetization is completely suppressed due to thermalization ($T > T_C$) at that position. Indeed, the plateau duration increases with $x$, consistent with the real-space observation that the stripe contrast reappears progressively from the left edge toward the right, as discussed above. In contrast, the relaxation after the plateau region shows a distinct behavior: $I_{FFT}$ at $t = 1600$ ns is significantly lower at $x = 700$ nm than at other $x$ positions, suggesting a delay in the recovery process only around this region, where the magnetic edge dislocation exists. The origin of this timescale will be discussed later.

To quantify these behaviors, we fit each time trace with a modified exponential recovery function, in which a simple exponential function with relaxation time $\tau$ and amplitude $A$ is truncated by a constant value at $0 < t < p$ [Figure 3(d)]. Here, $\tau$, $A$, and $p$ are treated as fitting parameters, and the intensity at $0 < t < p$ represents the background level when the magnetic contrast is completely suppressed. Exponential functions with the fitted relaxation time $\tau$ are shown as solid gray curves in Figure 3(c), and the values of plateau duration $p$ are plotted as black arrows. These analyses show that post-plateau recovery is well described by a single exponential function. When we plot the fitted $p$ and $\tau$ for each trace as a function of position $x$ [Figure 3(e)], these temporal parameters exhibit distinct behaviors: the plateau duration $p$ increases monotonically with the distance $x$ from the edge, while the relaxation time $\tau$ shows a clear maximum around $x = 700$ nm, coinciding with the position of the magnetic edge dislocation. This behavior contradicts the simple thermal diffusion model, since the monotonic temperature diffusion along the $x$ direction should result in a monotonic increase in both $p$ and $\tau$. The monotonic increase in $p$ is naturally explained by thermal diffusion toward the left edge as discussed above. In contrast, the non-monotonic change in $\tau$ suggests that the recovery of the magnetic stripe is delayed around the magnetic edge dislocation, as suggested above.

To further elucidate the characteristic relaxation dynamics around the magnetic edge dislocation discussed above, we analyze magnified LTEM images [Figure 4(a)] within the region highlighted by the solid rectangles in Figure 2(a). To visualize the contrast variations in the LTEM images, we separate the bright and dark regions using an adaptive threshold algorithm [Figure 4(b)]. For this analysis, the threshold value is set as the median value of neighboring 11 × 11 pixels (180 × 180 nm) added by 2.5 % of the median value of the entire image for clarity. Before excitation ($t < 0$) and at $t = 1500$ ns, a bright stripe terminates at the center of the image, indicating the magnetic edge dislocation. At $t = 800$ ns, however, segmentation around the magnetic edge dislocation becomes unclear due to pronounced blurring in the LTEM image. Since pump-probe LTEM measurements average over millions of independent events, the blurred contrast at $t = 800$ ns likely arise from the ensemble averaging of several different configurations. In other words, several competing relaxation paths are stochastically realized at $t = 800$ ns [Path A, B, and C in Figure 4(c)], and converging to one of them, the original state (Path B) at $t = 1500$ ns. In the case of Path A or C, the magnetic pattern should return to the original state after $t = 800$ ns via the slipping motion of the magnetic edge dislocation along the helical $q$-vector[20,21]. In these cases, the formation of the initial pattern (same as Path B) is delayed, which naturally explains the localized delay of relaxation around the magnetic edge dislocation. It should be noted that we cannot exclude the possibility of multiple slipping motions occurring within a single relaxation path, as such processes could also produce similar LTEM contrasts.

Finally, we discuss the spatiotemporal scale of the magnetic dynamics around the magnetic edge dislocation based on the experimental results described above. At the position of the magnetic edge dislocation, the relaxation is characterized by a timescale of approximately 1600 ns, whereas at positions away from the dislocation it is about 600 ns. Considering this difference, the stochastic slipping motion near the magnetic edge dislocation is likely to occur on a timescale of ~ 1 μs. Furthermore, the observed blurring in the LTEM images extends over roughly a few helical periods, suggesting that the influence of the edge dislocation emerges on a comparable spatial scale. Future investigations, particularly single-shot measurements capable of resolving individual relaxation pathways, will be important for further clarifying this characteristic spatiotemporal scale and for deepening the understanding of defect-mediated magnetic dynamics.

In summary, we investigated photothermally induced helical-to-paramagnetic phase transition in a chiral magnet $Co_9Zn_9Mn_2$ using pump-probe LTEM measurements. In addition to a directional magnetization recovery driven by thermal diffusion, we observed pronounced blurring of LTEM images around the magnetic edge dislocation. Fourier-transform analysis revealed a significant delay in magnetic contrast recovery at the magnetic edge dislocation. These distinct dynamics around the magnetic edge dislocation are well described by the stochastic ensemble average of several competing relaxation paths, involving the slipping motion of the magnetic edge dislocation. The present results may reveal the characteristic spatiotemporal scales for these dynamics and indicate that enhanced stochasticity may occur around topological defects during the recovery dynamics of magnetic phase transitions.

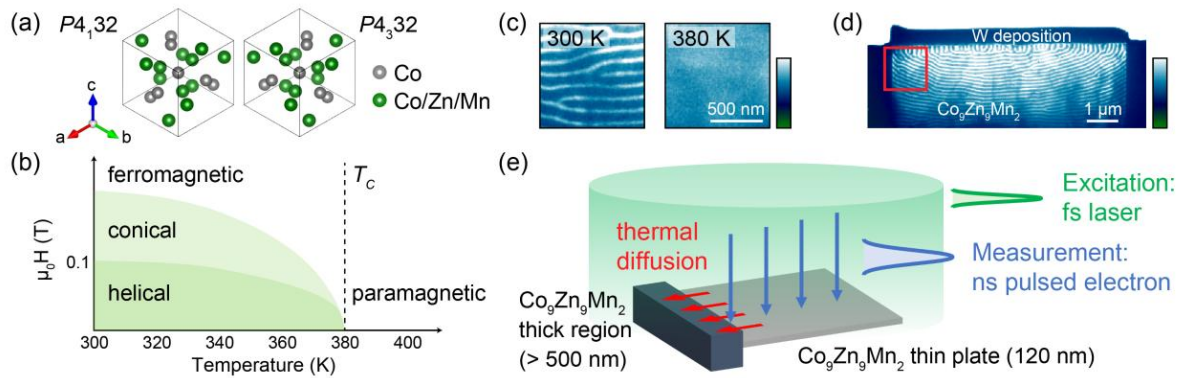

Figure 1. Sample characterization and experimental setup. (a) Crystal structure of $Co_9Zn_9Mn_2$. (b) Schematic magnetic phase diagram of the fabricated $Co_9Zn_9Mn_2$ thin plate. (c) Temperature-dependent LTEM images of the sample. (d) LTEM image of the entire $Co_9Zn_9Mn_2$ thin plate. Time-resolved LTEM measurements are performed in the region highlighted by the solid red rectangle. (e) Schematic experimental setup of the sample excited by the femtosecond laser.

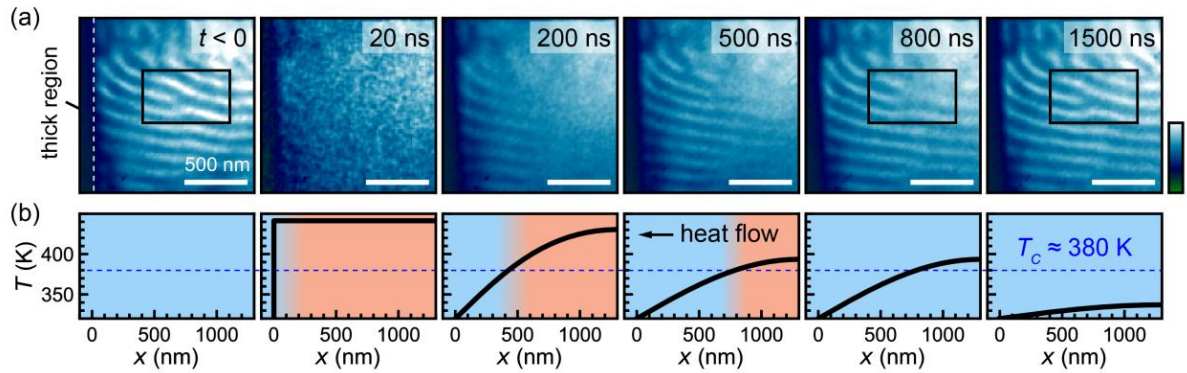

Figure 2. Photothermally induced helical-to-paramagnetic phase transition and recovery. (a) Time-dependent LTEM images of the sample. Solid rectangles highlight the magnetic edge dislocation. (b) Expected time evolution of temperature distribution along the $x$ axis. The Curie temperature is indicated as the dashed blue line.

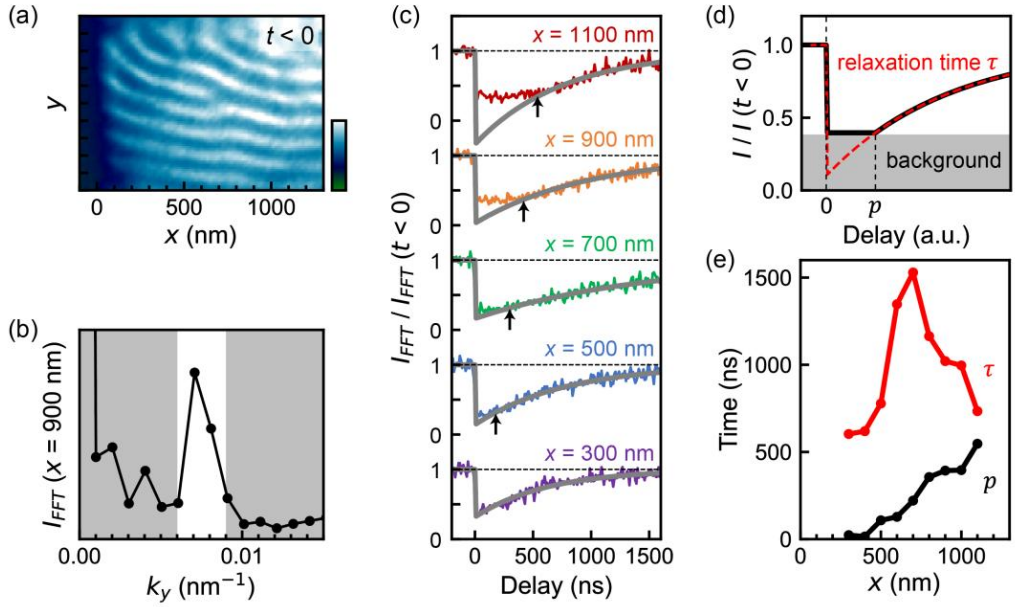

Figure 3. Quantitative analysis of magnetic contrast recovery. (a) LTEM image of the sample within the analysis region before laser excitation. (b) $x$-integrated one-dimensional FFT spectrum of Figure 3(a) along the $y$ axis. The unshaded region corresponds to the range of the helical wavenumber. (c) Time-dependent integrated FFT intensity at representative $x$ positions. Solid gray curves are exponential recovery functions with relaxation time $\tau$ obtained from fitting using Figure 3(d). Black arrows indicate the duration of the plateau. (d) Modified exponential recovery model with plateau duration $p$ and relaxation time $\tau$. (e) $x$ dependence of $p$ and $\tau$ obtained from fitting.

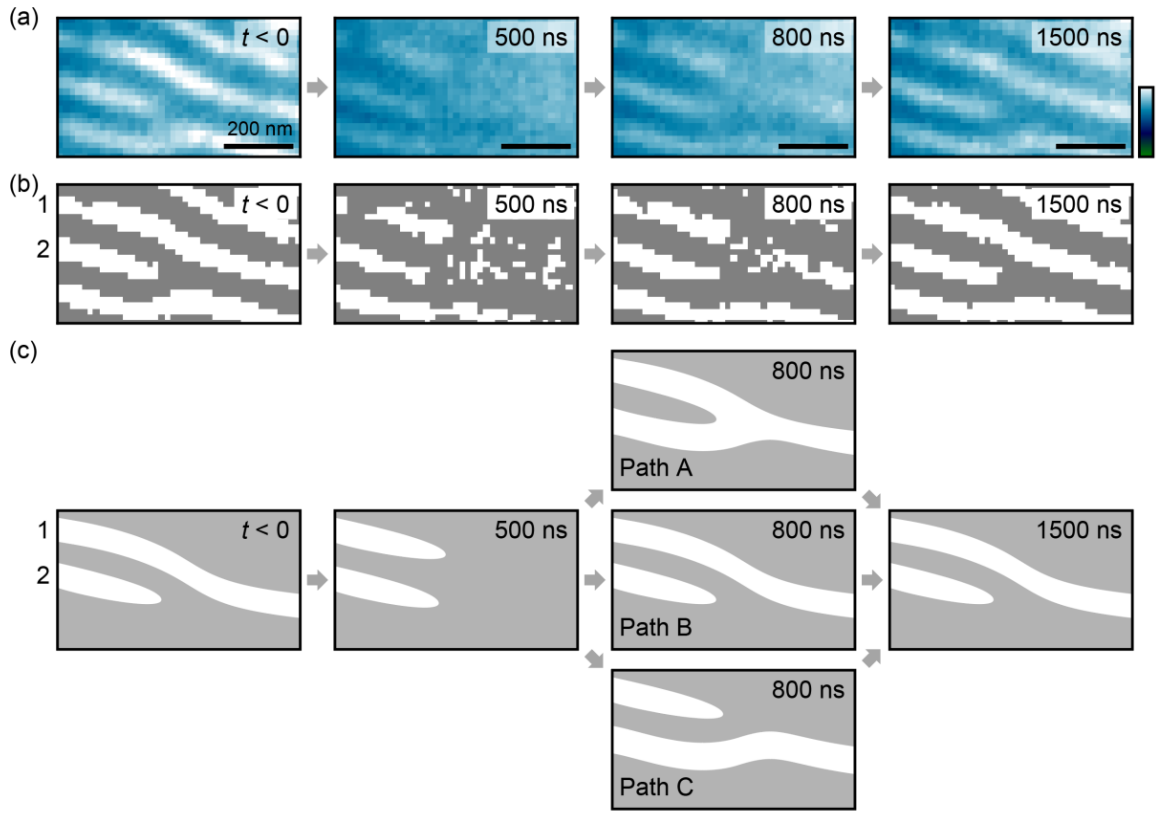

Figure 4. Detailed dynamics around the magnetic edge dislocation. (a) Magnified LTEM image around the magnetic edge dislocation at representative time delays. (b) Segmentation of bright and dark regions in Figure 4(a). (c) Schematic interpretation of the time evolution of the helical stripes labelled 1 and 2.

## Acknowledgements

This work was financially supported by the JST-CREST program (Grant No. JPMJCR20T1), JST-PRESTO program (Grant No. JPMJPR24JA), JSPS Grants-in-Aid for Scientific Research (KAKENHI) program (Grant No. 24H00410, No. 25K00057), Toray Science Foundation (Grant No. 23-6405), The Sumitomo Foundation and by the RIKEN TRIP initiative (Many-body Electron Systems and Advanced General Intelligence for Science).

## Conflict of Interest

The authors have no conflicts to disclose.

## Author Contributions

Dongxue Han: Conceptualization, Data curation, Formal analysis, Investigation, Methodology, Visualization, Writing – original draft

Asuka Nakamura: Conceptualization, Data curation, Formal analysis, Funding acquisition, Methodology, Project administration, Supervision, Visualization, Writing – review & editing

Takahiro Shimojima: Conceptualization, Funding acquisition, Methodology, Project administration, Resources, Supervision, Writing – review & editing

Kosuke Karube: Resources, Writing – review & editing

Yasujiro Taguchi: Resources, Writing – review & editing

Yoshinori Tokura: Resources, Funding acquisition, Writing – review & editing

Kyoko Ishizaka: Conceptualization, Funding acquisition, Project administration, Supervision, Writing – review & editing

## Data Availability

The data that support the findings of this study are available from the corresponding author upon reasonable request.